\documentstyle[eqsecnum,aps]{revtex}
\def\e{{\rm e}}
\def\aut#1{#1}

\def\cl{{\rm cl}}

\def\ins#1{}
\def\iem#1{{\em #1\/}}
\def\iems#1{}
\def\comment#1{}
\def\ind#1{#1}
\newcommand{\Tr}{{\rm Tr\,}}
 \newcommand{\tab}{_{t_a}^{t_b}}
\def\Det{{\rm  Det\,}}
\def\const{{\rm const\,}}

\def\aut#1{#1}

\def\cm#1{}

 \def\lfrac#1#2{{{{#1}/{#2}}}}

\begin{document}
\setcounter{figure}{0}
\Roman{figure}

\title{{
Fokker-Planck and Langevin Equations from
Forward--Backward Path Integral
}}
\author{Hagen Kleinert%
 \thanks{Email: kleinert@physik.fu-berlin.de 
URL:
http://www.physik.fu-berlin.de/\~{}kleinert \hfil
}}
\address{Institut f\"ur Theoretische Physik,\\
Freie Universit\"at Berlin, Arnimallee 14,
14195 Berlin, Germany}
\maketitle
\begin{abstract}
Starting from a
forward--backward path integral of a point particle
 in a bath of
harmonic  oscillators, we
derive the Fokker-Planck and Langevin equations with and without inertia.
Special emphasis is placed upon
the correct operator order in the time evolution operator.
The crucial step is the evaluation of a
Jacobian with a retarded time derivative
by analytic regularization.
\end{abstract}

%
\section{Introduction}
In 1963, Feynman and Vernon \cite{FV}
set up a path integral
for the study of a point particle
in a thermal bath of harmonic oscillators.
If the particle has a mass $M$, moves in a potential $V(x)$, and is coupled linearly
to
a large number of oscillators $X_i(t)$ of mass $M_i$ and frequency
$ \Omega _i$, the
probability to run
from
the spacetime point
 $x_at_a$ to
 $x_bt_b$ is given by
the forward--backward path
\begin{equation}
|( x_bt_b|x_at_a)|^2=\int
{\cal D}x_+
{\cal D}x_-\,\exp\left\{ \frac{i}{\hbar }
\int\tab\left[
\frac{M}{2}\left(x_+^2-x_-^2\right)
-V(x_+)+V(x_-)-x_+\sum_ic_iX^i_++x_-\sum_ic_iX^i_-\right] \right\},
\label{@PRO}\end{equation}
where $c_i$ are coupling strengths.
The bath oscillators are supposed to be
in thermal equilibrium at a temperature $T$.
This is taken into account by forming the
thermal average of the bath oscillators.
For a single oscillator,
this is done by
considering $c_ix_\pm$ as
external currents
$j_\pm$ coupled to $X_\pm$, and calculating
the Gaussian integral
\begin{eqnarray}
Z_0[j_+,j_-]= \int dX_b\,dX_a~(X_b\, \hbar  \beta |X_a0)_ \Omega  ~
( X_bt_b|X_at_a)_ \Omega^{j+}~
( X_bt_b|X_at_a)_ \Omega^{j_-*}.
\label{@ZPI}\end{eqnarray}
where $(X_b\, \hbar  \beta |X_a0)_ \Omega $ is the
imaginary-time amplitude
\begin{eqnarray} \!\!\!\!\!\!\!\!\!\!\!\!\!\!\!\!\!\!\!\!
(X_b \hbar  \beta \vert X_a0) &=&
 \frac{1}{     \sqrt{ 2\pi  \hbar/M }      }
   \sqrt{  \frac{\Omega }{\sinh  \hbar  \beta }       }
   \exp \left\{
-                \frac{1}{2\hbar}
		\frac{M\Omega }{\sinh  \hbar \beta  \Omega }
	    [(X_b^2 + X_a^2 )
		      \cosh \hbar  \beta \Omega
- 2X_b X_a]
    \right\}\label{2.130cX}
,
\end{eqnarray}
and
$( X_bt_b|X_at_a)_ \Omega^{j}
$
the path integral over the bath oscillator
\begin{eqnarray}\!\!\!\!\!\!\!\!\!\!( X_bt_b|X_at_a)_ \Omega^j
 &=&  \int {\cal D}X(t) \exp\left\{
         \frac{i}{\hbar} \int_{t_a}^{t_b} dt \left[
         \frac{M}{2} (\dot{X}^2 - \Omega ^2 X^2) + j X\right]\right\}
\nonumber \\&=&
e^{(i/\hbar ){\cal A}_{\cl,j}}F_{\Omega,j}(t_b,t_a).
\label{3.88X}
\end{eqnarray}
with a total classical action
\begin{eqnarray} \!\!\!\!\!\!\!\!\!\!\!\!\!\!\!\!\!\!\!\!\!\!\!\!\!\!{\cal A}_{\cl,j} &=&
\!\! \frac{1}{2}
           \frac{M\Omega }{\sin \Omega (t_b-t_a)}
           \!\left[ (X_b^2 + X_a^2) \cos \Omega  (t_b-t_a)
          \! -\! 2X_bX_a\right]\!  \nonumber  \\
 \!\!\!\!\!\!\!\!\!\!\!\!\!\!\!\!\!\!\!\!\!\!\!\!\!\!\!\!& +&\frac{1}{ \sin
             \Omega (t_b-t_a)}
             \int_{t_a}^{t_b} dt [X_a
             \sin \Omega  (t_b-t) + X_b \sin \Omega  (t-t_a)]j(t),\label{3.88aX}
\end{eqnarray}
and the fluctuation factor
\begin{eqnarray}
\!\!\!\!\!\!\!\!\!\!\!\!\!\!\!\!\!\!\!
   F_{\Omega,j}(t_b,t_a)&\!\!=\!\!&
  \frac{1}{\sqrt{2\pi  i\hbar/M}} \sqrt{\frac{\Omega }
       {\sin \Omega (t_b-t_a)}}  \nonumber \\
\!\!\!\!\!\!\!\!\!\!\!\!\!\!\!\!\!\!
\!\!\!\!\!\!\!\!\!\!\!\!\!\!\!\!
&\times& \exp \left\{
-\frac{i}{\hbar M\Omega  \sin \Omega (t_b-t_a)}
            \int_{t_a}^{t_b}\!dt  \int_{t_a}^{t}\! dt'
            \sin \Omega (t_b-t) \sin \Omega  (t'-t_a) j(t) j (t')
 \right\}\!. \label{3.88fl0}
\end{eqnarray}
The result of the thermal average
in Eq.~(\ref{@ZPI}) is
\begin{eqnarray}
&&\!\hspace{-1cm}Z_0[j_+, j_- ]  = \exp\bigg\{
  -\frac{1}{2\hbar } \int dt \int dt'\Theta(t-t')  \label{18.148bt0}
 \\
&&\!\!\times  \Big[  (j_+ -j_-)(t)G^R(t,t')(j_+ +j_-)(t')
               +
  (j_+ -j_-)(t)A(t,t')(j_+ -j_-)(t')
  \Big],  \bigg\}\nonumber
\end{eqnarray}
where $A(t,t')$ and $G^R(t,t')$
are the expectation values
$\langle \{\hat X(t),\hat X(t')\}\rangle $
and $\Theta(t-t')\langle \hat [\hat X(t),\hat X(t')]\rangle $
in operator language.
The are real and imaginary parts of the time-ordered Green function
and $G(t,t')=\Theta(t-t')\langle \hat T\hat X(t),\hat X(t')\rangle $:
\begin{equation}
G(t,t')= A(t,t')+
 G^R(t,t')=\frac{\hbar }{2M \Omega }
\frac{\displaystyle\cosh\frac{ \Omega}{2}\left[\hbar  \beta -i(t-t')
\right]}
{\displaystyle\sinh\frac{\hbar  \Omega \beta }{2}}
,~~~t>t'..
\label{@expo in}\end{equation}
which is the analytic continuation
of the periodic imaginary-time Green function
to $\tau =it$.

The thermal average of the probability (\ref{@PRO})
is then given by the forward--backward path integral
\begin{eqnarray}
\!\!\!\!\!\!\!\!\!\!\!\!\!\!\!\!\!\!\!\!|({x}_b t_b|{x}_a t_a)|^2  &=&
  \int {\cal D}x_+(t)\int {\cal D}x_-(t)\,
  \nonumber \\
\!\!\!\!\!\!\!\!\!\!\!\!\!\!\!\!\!\!\!\!\! &\times&     \exp\left\{
 \frac{i}{\hbar}\int\tab dt\,
    \left[ \frac{M}{2}({\dot x_+}^2-{\dot x_-}^2)-
    (V(x_+)-V(x_-))\right] +\frac{i}{\hbar }
    {\cal A}^{\rm FV}[x_+,x_-]\right\}.
 \label{18.389n}
\end{eqnarray}
where $\exp{i{\cal A}^{\rm FV}[x_+,x_-]/\hbar }$ is the Feynman-Vernon
{\em influence functional\/} defined by
\begin{eqnarray}
&&\!\hspace{-.6cm}Z_0^{\rm b}[x_+, x_- ]\equiv
\exp\left\{i{\cal A}^{\rm FV}[x_+,x_-]/\hbar  \right\}\equiv
\exp\left\{
i{\cal A}^{\rm FV}_D[x_+,x_-]/\hbar
+i{\cal A}^{\rm FV}_F[x_+,x_-]/\hbar
\right\}
\label{18.148bta}
 \\
&= &
 \exp\bigg\{
  -\frac{1}{2\hbar  } \int dt \int dt'\,\Theta(t-t')  \label{18.148bt}
 \Big[  ( x_+ - x_-)(t)G^{A}_{\rm b}(t,t')( x_+ + x_-)(t')
               +
  ( x_+ - x_-)(t)A_{\rm b}(t,t')( x_+ - x_-)(t')
  \Big]  \bigg\}.       \nonumber
\end{eqnarray}
with
$G^{A}_{\rm b}(t,t')$ and $A_{\rm b}(t,t')$ being
commutator and anticommutator
functions of the bath.
The first  and second parts of the exponents have been
distinguished as dissipative and fluctuating parts
${\cal A}^{\rm FV}_D[x_+,x_-]$ and
${\cal A}^{\rm FV}_F[x_+,x_-]$
of
 of the effective influence
action
${\cal A}^{\rm FV}[x_+,x_-]$.

The bath functions
$G^{A}_{\rm b}(t,t')$ and $A_{\rm b}(t,t')$
 are
sums of correlation functions
of the individual oscillators of mass $M_i$ frequency $ \Omega _i$, each contributing
with a weight $c_i^2$.
 Thus we may
write
\begin{eqnarray} \label{@}
\hspace{-.9cm}G^A_{ \rm b}(t,t')&=&-\Theta(t'-t)    \sum _ic_i^2
\langle [\hat X_i(t),\hat X_i(t')]\rangle_ \rho~~=
 - \hbar \int _{-\infty}^\infty \frac{d \omega'}{2\pi}
\sigma_{\rm b} ( \omega' )i\sin \omega'(t-t'),\nonumber \\
\hspace{-.9cm}A_{\rm b}(t,t' )&=&  \sum _ic_i^2
\langle\left\{\hat X_i(t),\hat X_i(t')\right\}\rangle_ \rho\!=\!
  \hbar \int _{-\infty}^\infty \frac{d \omega}{2\pi}\sigma_{\rm b} ( \omega' )
 \coth\frac{\hbar  \omega' }{2k_BT}~\cos \omega'(t-t')
 ,   \label{18.386}
\end{eqnarray}
where $\sigma_{\rm b}(\omega') $  is the spectral density of the bath
\begin{equation}
\sigma_{\rm b}(\omega') \equiv  2\pi \sum_i\frac{ c_i^2  }{2M_i\Omega_i}
 [ \delta(\omega '-\Omega_i)-\delta(\omega' +\Omega_i)]
  \label{18.387}
\end{equation}

For a discussion of the properties  of the influence functional,
we introduce an auxiliary \ind{retarded function}
\begin{equation}
 \gamma(t-t')\equiv \Theta(t-t')\frac{1}{ M}\int _{-\infty}^\infty\frac{d \omega}{2\pi}\frac{ \sigma( \omega)}{ \omega}e^{-i \omega(t-t')},
\label{18.gammafu}\end{equation}
and write
\begin{equation}
G^R_{\rm b}(t,t')=i{\hbar }{M} \dot  \gamma(t-t')+i \Delta \omega^2 \delta(t-t')
\label{18.Cdeco}\end{equation}
with
\begin{equation}
\!\!\!\!\!\!\!\!\!{M}\Delta \omega ^2\equiv
-\int _{-\infty }^{\infty}\frac{ d\omega'}{2\pi }\frac{\sigma _{\rm b} (\omega ')}{\omega '}
=- \sum _i\frac{c_i^2}{M_i\omega _i^2}.
\label{18.deltaome}\end{equation}
  Inserting the first term in the decomposition (\ref{18.Cdeco}) into
(\ref{18.148bt}), the dissipative part of the influence
functional can be integrated by parts in $t'$ and becomes
\begin{eqnarray}
\!\!\!\!\!
\!\!\!\!\!\!\!\!\!\!\!\!\!
{\cal A}_D^{\rm FV}[x_+,x_-]&=&
-\frac{M}{2}\int_{t_a}^{t_b} dt\int_{t_a}^{t_b} dt'\,
(x_+-x_-)(t) \gamma(t'-t)
(\dot x_++\dot x_-)(t')
\nonumber  \\
&&
+\frac{M}{2} \int_{t_a}^{t_b} dt (x_+-x_-)(t)
\gamma(t_b-t)(x_++x_-)(t_a)
\!.
\label{18.inflfC}\end{eqnarray}
The $ \delta$-function in  (\ref{18.Cdeco})
contributes
to $ {\cal A}_C^{\rm FV}[x_+,x_-]$
a
term
\begin{eqnarray}
 \Delta {\cal A}_{\rm loc}[x_+,x_-]=
\frac{M}{2} \int_{t_a}^{t_b} dt
\Delta \omega^2  (x_+^2-x_-^2)(t)
,
\label{18.inflfC3}
\end{eqnarray}
which may simply be absorbed into
the potential terms in the path integral
(\ref{18.389n}) renormalizing them to
\begin{eqnarray}
- \frac{i}{\hbar}\int\tab dt\,
    \left[
    V_{\rm ren}(x_+)-V_{\rm ren}(x_-)
\right],
 \label{18.389nn}
\end{eqnarray}

The odd bath function
$ \sigma_{\rm b} ( \omega ')$ can be expanded
in a power series with only odd powers of $\omega'$. The lowest
approximation
\begin{equation} \label{x18.199}
\sigma_{\rm b}(\omega')\approx 2 M\gamma \omega',
\end{equation}
describes Ohmic dissipation
with some friction constant $\gamma $.
For frequencies much larger than the atomic relaxation rates,
the friction goes to zero.
This behavior is modeled by the \ind{Drude form}
of the
spectral function
\begin{equation}
\sigma (\omega' )\approx 2M\gamma  \omega'\frac{ \omega _D^2}{\omega _D^2+\omega'{} ^2}.
\label{18.3.drfor}\end{equation}
Inserting this into Eq.~(\ref{18.gammafu}),
we obtain the Drude form of the function
$ \gamma(t)$:
\begin{equation}
 \gamma_D^R(t)\equiv \Theta(t) \, \gamma \omega_De^{- \omega_D|t|}.
\label{18.gammafuD}\end{equation}
The superscript emphasizes the
retarded nature.
This can also be written as a Fourier integral
\begin{equation}
 \gamma_D(t)=\int _{-i\infty}^\infty \frac{d \omega}{2\pi}
 \gamma_D( \omega)e^{-i \omega t},
\label{18.gammafuDt}\end{equation}
with the Fourier components
\begin{equation}
 \gamma_D^R( \omega')= \gamma\frac{ i\omega_D}{ \omega'+ i\omega_D}.
\label{@Drom}\end{equation}
The position of the pole in the lower half-plane
 ensures the
retarded nature of the friction term by
producing the Heaviside function  in (\ref{18.gammafuD}).

In the Ohmic limit
(\ref{x18.199}),
the dissipative part
 of the influence functional simplifies.
Then $ \gamma(t)$ becomes narrowly peaked at positive $t$
may be expressed in terms
of a
{\em left-sided\/} $\delta$-function%
\iems{right-sided $\delta$-function}%
\iems{one-sided $\delta$-function}
as
\begin{equation}
 \gamma(t)\rightarrow
2 \gamma \,\delta^R(t),
\label{@}\end{equation}
whose  superscript $R$ indicates the
retarded asymmetry of the $ \delta $-function,
which
has the property that
\begin{equation}
\int dt \Theta(t)  \delta ^R(t)=1.
\label{@}\end{equation}
With this,  (\ref{18.inflfC}) becomes a local action
\begin{eqnarray}
\!\!\!\!\!\!\!\!\!\!\!{\cal A}_D^{\rm FV}[x_+,x_-]=
-\frac{M}{2} \gamma\int_{t_a}^{t_b}\! dt\!
\,(x_+-x_-)
(\dot x_++\dot x_-)^R
-\frac{M}{2}  \gamma (x_+^2-x_-^2)(t_a).
\label{18.inflfCl}\end{eqnarray}
The right-sided nature of the $ \delta $-function
causes  an infinitesimal {\em negative\/} shift
in  the time argument of
 the velocities
 $(\dot x_++\dot x_-)(t)$
with respect to the factor $(x_+-x_-)(t)$,
indicated by the superscript $R$.
 It
expresses the \iem{causality}
of the friction forces
and will be seen to be crucial in producing a
probability conserving
time evolution of the probability
distribution.

The second term changes only the curvature of the effective potential
at the initial time, and
can be ignored.
In the first term it is important to observe that
the retarded nature of the dissipative term
and of the function $ \gamma (t)$ in (\ref{18.gammafu})
ensures
that the velocity term $(\dot x_++\dot x_-)(t)$ lies {\em before\/}
 $(x_+-x_-)(t)$
in a time-sliced path integral.
This ensures the \iem{causality}
of the friction forces.

It is useful to incorporate the slope information
(\ref{x18.199}) also into the
 bath correlation function $A_{\rm b}(t,t')$
in (\ref{18.386}), and factorize it as
\begin{eqnarray}
A_{\rm b}(t,t' )=2M \gamma k_BTK(t,t')
 ,   \label{18.386k}
\end{eqnarray}
where
\begin{eqnarray}
K(t,t')=K(t-t')&\equiv&
\frac{1}{2M \gamma k_BT}  \sum _ic_i^2
\langle\{\hat  X_i(t),\hat  X_i(t')\}\rangle_ \rho~\nonumber \\
&=&
  \int _{-\infty}^\infty \frac{d \omega'}{2\pi}
K( \omega')
e^{-i \omega'(t-t')}
 ,   \label{18.390}
\end{eqnarray}
with Fourier transform
\begin{eqnarray}
K( \omega')\equiv
\frac{1}{2M \gamma }
\frac{\sigma ( \omega' )}{\omega'}
\frac{\hbar  \omega' }{2k_BT} \coth\frac{\hbar  \omega' }{2k_BT},
    \label{18.386k3}
\end{eqnarray}
which in the limit of a purely  \ind{Ohmic dissipation} simplifies to
\begin{eqnarray}
K( \omega')=K^{\rm Ohm}( \omega')\equiv
\frac{\hbar  \omega' }{2k_BT} \coth\frac{\hbar  \omega' }{2k_BT}.
    \label{18.391}
\end{eqnarray}
The function $K( \omega')$
has
the
normalization $K(0)=1$,
giving $K(t-t')$
a unit temporal area:
\begin{equation}
\int_{-\infty}^{\infty} dt\, K(t-t') = 1.
  \label{18.392}
\end{equation}
In the classical limit $\hbar \rightarrow 0$,
\begin{equation}
K( \omega')=\frac{ \omega_D^2}{ \omega'^2+ \omega_D^2},
\label{@}\end{equation}
and
\begin{equation}
K(t-t')=\frac{1}{2 \omega_D}e^{- \omega_D(t-t')}.
\label{@DRUDE}\end{equation}
In the limit of Ohmic dissipation, this becomes a $ \delta$-function.
Thus $K(t-t')$ may be viewed as a $ \delta$-function broadened
by quantum  fluctuations and relaxation effects.

With the function $K(t,t')$, the fluctuation part of the influence functional
in
(\ref{18.148bta}),
(\ref{18.148bt}),
(\ref{18.389n}) becomes
\begin{equation}
{\cal A}_F^{\rm FV}[x_+,x_-]=
i  \frac{M\gamma k_B T}{\hbar}
    \int_{t_a}^{t_b} dt\, \int_{t_a}^{t_b} dt'\,
    (x_+ -x_-)(t)\,K(t,t')\,(x_+ -x_-)(t').
\label{18.inflfA}\end{equation}
Here we have used the symmetry of the function
$K(t,t')$ to remove the Heaviside function
$\Theta(t-t')$ from the integrand, extending the range of $t'$-integration
to the entire interval $(t_a,t_b)$.

In the Ohmic limit, the
probability of the particle to
move from ${x}_a t_a$ to
${x}_b \,t_b$ is given by the path integral
\begin{eqnarray}  \!\!\!\!\! \!\!\!\!\!\!\!\!\!\!\!
|({x}_b t_b|{x}_a t_a)|^2  &=&
  \int {\cal D}x_+(t)\int {\cal D}x_-(t)\,
  \nonumber \\
&\times &     \exp\left\{
 \frac{i}{\hbar}\int\tab dt\,
    \left[ \frac{M}{2}({\dot x_+}^2-{\dot x_-}^2)-
    (V(x_+)-V(x_-))\right]
    \right\} \nonumber \\
&\times&
\exp\left\{ -i\int\tab dt\,
  \frac{M\gamma}{2\hbar}(x_+ -x_-)(t)(\dot x_+ +\dot x_-)^R(t)
\right .\nonumber  \\   &&\,\!\!\!\!\left .~~
~~~~~~~-
  \frac{M\gamma k_B T}{\hbar^2}
    \int_{t_a}^{t_b} dt\, \int_{t_a}^{t_b} dt'\,
    (x_+ -x_-)(t)\,K(t,t')\,(x_+ -x_-)(t')
  \right\}.
\label{18.389}
\end{eqnarray}
This
is the \iem{closed-time path integral\/}
of a particle in contact with a thermal reservoir.

The paths $x_+(t), x_-(t)$
may also be associated with a forward and a backward movement of the particle in time.
For this reason, (\ref{18.389}) is also called a \iem{forward--backward path integral}.
The
hyphen is pronounced as  {\em minus\/},
to emphasize the opposite signs in the partial actions.

It is now convenient to change integration variables
and go over to average and relative
coordinates of the two paths $x_+$, $x_-$:
\begin{eqnarray}
x & \equiv  & (x_ + +x_-)/2, \nonumber \\
y&\equiv & x_+ -x_-.
  \label{18.393}
\end{eqnarray}
Then (\ref{18.389}) becomes
\begin{eqnarray}
|({x}_b t_b|{x}_a t_a)|^2 &=&
  \int {\cal D}x(t)\int  {\cal D}y(t)\,
 \exp\bigg\{
-\frac{i}{\hbar}
    \int \tab dt\, \left[M \left(-\dot y\dot x+\gamma y\dot x^R\right)+
V\left(x+\frac{y}{2}\right)-V\left(x-\frac{y}{2}\right)\right]
    \nonumber \\
&   &\!\,\!\!\!\!\!\!\!\!~~~~~~~~~~~~~~~~~~~~~~~~~~~~~~~~~~~- \,
    \frac{M\gamma k_B T}{\hbar^2}
    \int\tab dt\, \int_{t_a}^{t_b} dt'\,
      y(t)K(t,t')y(t') \bigg\}.
  \label{18.394}
\end{eqnarray}

\section[Fokker-Planck Equation]
{Fokker-Planck Equation\label{sec18.6}}
\index{Fokker-Planck equation}

At high-temperatures,
the Fourier transform of the Kernel $K(t,t')$
in Eq.~(\ref{18.391}) tends to unity such that
$K(t,t') $ becomes a  $\delta $-function,
and  the bath correlation function (\ref{18.386k})
becomes approximately
\begin{eqnarray}
A_{\rm b}(t,t' )\approx \frac{1}{\hbar } w \, \delta (t-t'),
    \label{18.386kdel}
\end{eqnarray}
where we have introduced the constant
proportional to the temperature:
\begin{equation}
w \equiv 2M\gamma k_BT,
  \label{18.400}
\end{equation}
which is  related to the
so-called  \iem{diffusion constant}
\begin{equation} \label{@}
D\equiv k_B T/M\gamma
\end{equation}
by
\begin{equation} \label{@wDel}
w = 2 \gamma ^2M^2D/T.
\end{equation}

Then the  path integral
 (\ref{18.394})  for the probability distribution of a particle
coupled to a thermal bath
simplifies
to
\begin{eqnarray} \label{@}
\lefteqn{\!\!\!\!\!\!\!\!\!\!
P(x_b t_b|x_a t_a) \equiv
  |(x_b t_b|x_a t_a)|^2 =
  \int {\cal D}x(t)\, \int {\cal D}y(t)\,
  }\nonumber \\
&&\times     \exp\left\{ -\frac{i}{\hbar}\int_{t_a}^{t_b} dt\,
    y[M\ddot x+M\gamma\dot x^R+V'(x)]-
    \frac{ w }{2\hbar^2}\int_{t_a}^{t_b} dt\, y^2\right\} .
  \label{18.410}
\end{eqnarray}
The superscript $ R $ records the
infinitesimal backward shift of the time argument
as in Eq.~(\ref{18.inflfCl}).
The
$y$-variable
can be integrated out,
and we obtain
\begin{eqnarray}
P(x_b t_b|x_a t_a)
 =  \int {\cal D}x(t)\, \exp\left\{ -\frac{1}{2  w }
  \int_{t_a}^{t_b} dt\, [M\ddot x+M\gamma\dot x^R+V'(x)]^2\right\}\!.
  \label{18.411}
\end{eqnarray}
This looks like a euclidean path integral associated
with the Lagrangian
 \begin{equation}
L_{\rm e} = \frac{1}{2 w } [M\ddot x+M\gamma\dot x+V'(x)]^2.
  \label{18.412}
\end{equation}
The solution of such path integrals with squares of second time derivatives
in the Lagrangian is given in Ref.~\cite{ddot}.
The result will, however, be different,
due to time-ordering of the $\dot x^R$-term.

Apart from this,
the Lagrangian
is not of the
conventional type since it involves a second time derivative.
The action principle $ \delta {\cal A}=0$  now
yields the \ind{Euler-Lagrange equation}
\begin{equation}
\frac{\partial L}{\partial x}-
\frac{d}{dt}\frac{\partial L}{\partial \dot x}
+\frac{d^2}{dt^2}\frac{\partial L}{\partial \ddot x}=0.
\label{@}\end{equation}
This equation can also be derived
via the usual \ind{Lagrange formalism}
by
considering $x$ and $\dot x$ as
independent generalized coordinates $x$, $v$.
\comment{
and
rewriting the Lagrangian
as $L_\e(x,\dot x,\ddot x)=L_\e(x,v,\dot v)+ ip (\dot x-v)$,
with some Lagrange multiplier $ p $.}

\section{Canonical Path Integral for Probability Distribution}
It is well-known that a
path integral
satisfies
a Schr\"odinger type of equation.
For the path integral
(\ref{18.411}) this
is known as a
\iem{Fokker-Planck equation}.
The relation is established (see
the textbook  \cite{PI})
by rewriting the path integral in canonical form.
Treating $v=\dot x$ as an independent dynamical variable,
 the canonical momenta of $x$ and $v$
are (see
Section~17.3 of the textbook
 Ref.\cite{GFCM})
\begin{eqnarray}
p  &= & i\frac{\partial L}{\partial \dot x} =
  i\frac{M \gamma }{ w }[M\ddot x+M\gamma\dot x+V'(x)]\nonumber \\
  &&~~~~~~\hspace{1pt}=
  i\frac{M \gamma }{ w }[M\dot v+M\gamma v+V'(x)],
  \nonumber \\
p_v & = & i\frac{\partial L}{\partial \ddot x} =
 \frac{1}{ \gamma }p.
  \label{18.418}
\end{eqnarray}
The Hamiltonian is given by the Legendre transform
\begin{eqnarray}
\hspace{-1cm}H(p,p_v,x,v) &=&
  L_{\rm e}(\dot x,\ddot x)-
  \sum_{i=1}^2 \frac{\partial L_{\rm e}}{\partial \dot x_i}\dot x_i
  \\
\hspace{-1cm}& = &
  L_{\rm e}(v,\dot v)+ipv+ip_v\dot v,
  \label{18.419}
\end{eqnarray}
where $\dot v$ has to be eliminated in favor of $p_v$ using
(\ref{18.418}).
This leads to
\begin{equation}
H(p,p_v,x,v) = \frac{  w }{2M^2}{p_v^2} -
  ip_v[ \gamma v+\frac{1}{M}V'(x)] + ip v.
  \label{18.420}
\end{equation}
The
the canonical path integral representation for the
probability reads therefore
\begin{eqnarray} \label{@}
\lefteqn{ \hspace{-1cm}
P(x_b t_b|x_a t_a) =
  \int {\cal D}x\, \int \frac{{\cal D}p}{2\pi}\, \int{\cal  D}v\, \int
  \frac{{\cal D}p_v}{2\pi}\,}\nonumber \\
  &&~~~~~~\times
  \exp\bigg\{ \int_{t_a}^{t_b}dt\,
    \left[i(p\dot x+p_v \dot v)-H(p,p_v,x,v)\right]
    \bigg\}.
  \label{18.421}
\end{eqnarray}
It is easy to verify that the path integral over $p$
enforces $v\equiv \dot x\equiv \dot x$, after which
the path integral
over $p_v$ leads back to the initial expression (\ref{18.411}).
We may keep the auxiliary variable $v(t)$ as an independent fluctuating
quantity in all formulas and decompose
the probability $P(x_b t_b|x_a t_a)$
with respect to the content of $v$ as an integral
\begin{equation}
P(x_b t_b|x_a t_a) =
\int_{-\infty}^{\infty}  dv_b\, \int_{-\infty}^{\infty}
 dv_a\, P(x_b v_b t_b|x_a v_a t_a).
  \label{18.422}
\end{equation}
The more detailed probability
on the right-hand side
has the path integral representation
\begin{eqnarray}
&&\!\!\!\!\!\!\!\!\!\!\!\!\!\!\!\!\!\!\!\!
P(x_b v_b t_b|x_a v_a t_a)  =
  |(x_b v_b t_b|x_a v_a t_a)|^2
 =
  \int {\cal D}x\, \int \frac{{\cal D}p}{2\pi}\, \int {\cal D}v\, \int \frac{{\cal D}p_v}{2\pi}\,
\nonumber \\
&&~~~~~~~~~~~~~~\times
  \exp\left\{ \int_{t_a}^{t_b} dt\,
    [i(p\dot x+p_v\dot v)-H(p,p_v,x,v)] \right\},
  \label{18.423}
\end{eqnarray}
where the end points of $v$ are now kept fixed at $v_b = v(t_b)$,
$v_a = v(t_a)$.

We now use the relation between a canonical path integral
and the Schr\"odinger equation
to conclude
that the probability distribution  (\ref{18.423})
satisfies the
Schr\"odinger-like
differential equation:\footnote{See the review paper by
\aut{S.~Chandrasekhar},
Rev.~Mod.~Phys.~{\em 15}, 1 (1943).}
\begin{equation}
H(\hat p,\hat p_v,x,v)P(x\, v\, t_b|x_a v_a t_a) =
  -\partial_{t} P(x\, v\, t|x_a v_a t_a).
  \label{18.424}
\end{equation}
This is the\index{Fokker-Planck equation}
{\iem{Fokker-Planck equation in the presence of
inertial forces}}.

At this place we note that when going over from the classical Hamiltonian
(\ref{18.420}) to the Hamiltonian operator
in the differential equation
(\ref{18.424}) there is an operator ordering problem.
When writing down Eq.~(\ref{18.424})
we do not
know in which order
the momentum $p_v$ must stand with respect
to $v$.
If we were dealing with an ordinary functional integral
in (\ref{18.411}) we would know the order.
It would be found as in the case of the
electromagnetic interaction
to by symmetric:
$-(\hat p_v \hat v+
 \hat v
\hat p_v )/2$.

On physical grounds, it is easy to guess
the correct order
The differential equation (\ref{18.424})
has to conserve the total probability
$\int dx\,dv P(x\, v\, t_b|x_a v_a t_a) $ for all times $t$.
This is
guaranteed
if all momenta stand to the left of
all coordinates in the
Hamiltonian operator.
Indeed, integrating the Fokker-Planck equation
(\ref{18.424})
over $x$ and $v$,
only a  left-hand position of the momentum operators
leads to a vanishing integral, and thus to a time independent total probability.
We suspect that this order must be derivable
from the retarded nature
of the velocity in the term $y\dot x^R$ in
(\ref{18.410}).
The proof that this is so is the essential point of this
paper, by which it goes beyond an earlier treatment of this subject in Ref.~\cite{SCHM}.

\section{Solving the Operator Ordering Problem}
\label{@OPOPR}
Since the ordering problem
in the Hamiltonian operator associated with (\ref{18.420})
does not involve the potential $V(x)$, we study  this problem
most simply
by considering the free Hamiltonian
\begin{equation}
H_0(p,p_v,x,v) = \frac{  w }{2M^2}{p_v^2} -
  i \gamma p_v  v + ip v.
  \label{18.420fr}
\end{equation}
which is associated with the Lagrangian path integral
\begin{eqnarray}
P_0(x_b t_b|x_a t_a)
 \propto  \int {\cal D}x(t)\, \exp\left\{ -\frac{1}{2  w }
  \int_{t_a}^{t_b} dt\, [M\ddot x+M\gamma\dot x^R]^2\right\}\!.
  \label{18.411fr}
\end{eqnarray}
We furthermore may concentrate  on the
probability with $x_b=x_a=0$,
and assume $t_b-t_a$ to be very large.
Then the frequencies of all Fourier decompositions
are continuous.

Forgetting for a moment the retarded nature of the velocity
 $\dot x$, the
Gaussian path integral can immediately be done
and yields
\begin{eqnarray}
P_0(0\, t_b|0\, t_a)
& \propto& \Det ^{-1}(- \partial _t^2
 -\gamma\partial_t)
\nonumber \\
&\propto& \exp\left[-(t_b-t_a) \int _{-\infty}^\infty\frac{d \omega }{2\pi}
\log ( \omega'{} ^2-i \gamma  \omega' )\right] .
  \label{18.411fr1c}
\end{eqnarray}
The integral can be evaluated in analytic regularization according to the rule
\begin{equation}
\int _{-\infty}^\infty\frac{d \omega' }{2\pi}
\log( \omega'\pm i  \gamma ) =  \frac{|  \gamma|  }{2},
\label{@intanal1}\end{equation}
which follows directly by integrating
in $ \gamma $
the symmetric decomposition of the integral
$\int d \omega '/( \omega '^2+ \gamma ^2)=\pi/| \gamma |$.
Hence we obtain for
the functional determinant in
(\ref{18.411fr1c}):
\begin{eqnarray}
\Det (- \partial _t^2 -\gamma\partial_t)&=&
\Det (i \partial _t)\Det(i\partial _t -i\gamma)=
\exp\left[ \Tr\log(i\partial _t)
+
\Tr\log(i\partial _t-i\gamma)\right]
\nonumber \\
&=&\exp\left[(t_b-t_a) \frac{\gamma}2 \right]
\label{@fudetg1}\end{eqnarray}
and thus
\begin{eqnarray}
\!\!\!\!\!\!\!\!\!\!\!\!\!\!\!\!
P_0(0\, t_b|0\, t_a)
& \propto&  \exp\left[ -(t_b-t_a) \frac{\gamma}2\right]  ,
  \label{18.411fr1x}
\end{eqnarray}
This corresponds to an energy $ \gamma /2$ and an
ordering $-i \gamma (\hat p_v v+v\hat p_v)/2$
in the Hamilton operator.

We now take the retarded time argument of $\dot x^R$ into account.
Specifically, we replace the term
$ \gamma y\dot x^R$ in (\ref{18.411fr})
by
$\int dtdt'\,y \,\gamma_D^R(t-t')\, x(t)$ containing
 the retarded \ind{Drude function}\ins{Drude friction}\ins{friction, Drude form}
(\ref{18.gammafuD})
of the friction.
The the frequency integral in
(\ref{18.411fr1c}) becomes
\begin{equation}
 \!\int _{-\infty}^\infty\frac{d \omega }{2\pi}
\log \left( \omega'{} ^2- \gamma  \frac{\omega' \omega _D}{ \omega'\! +\!i \omega _D} \right)
\!=\! \int _{-\infty}^\infty\frac{d \omega }{2\pi}
\left[ -\!\log ( \omega '\!+\!i \omega _D)
+\log\left( \omega'{} ^2\!+\!i  \omega _D\!- \!\gamma  \omega' \omega _D
 \right)\right]  ,
\label{@}\end{equation}
where we have omitted a vanishing integral over $\log \omega '$
by (\ref{@intanal1}).
We now use
(\ref{@intanal1})
to find
\begin{equation}
 \int _{-\infty}^\infty\frac{d \omega }{2\pi}
\left[ -\log ( \omega '+i \omega_D)
+\log\left( \omega'{} ^2+i \omega '  \omega _D- \gamma  \omega _D
 \right)\right] =
-\frac{ |\omega _D|}{2}
+\frac{|\omega _1|}{2}
+\frac{ |\omega _2|}{2} ,
\label{@tracelogD}\end{equation}
where $-i \omega _{12}$ are the solutions of
the quadratic equation
\begin{equation}
  \omega'{} ^2+i  \omega _D- \gamma  \omega' \omega _D=0.
\label{@}\end{equation}
For a large Drude frequency $ \omega _D$, they are given by
\begin{equation}
 \omega _1= \omega _D- \gamma ,~~~~~~ \omega _2= \gamma .
\label{@}\end{equation}
Inserting these into (\ref{@tracelogD})
we find a vanishing integral
 rather than $ \gamma $
in (\ref{18.411fr1x}), and thus a
functional determinant
\begin{equation}
\Det (- \partial _t^2 -\gamma\partial_t^R) =
\exp\left[ \Tr\log(- \partial _t^2 -\gamma\partial_t^R)\right]
=1,
\label{@fudetg1r}\end{equation}
instead of (\ref{@fudetg1}).
The notation $ \gamma \partial _{t}^R$
symbolizes now
 the specific retarded functional matrix a la Drude
with a large $ \omega _D$:
\begin{equation}
 \gamma \partial_t^R(t,t')\equiv \int dt''  \gamma^R _D(t-t'')
\partial _{t''}  \delta (t''-t').
\label{@retardedfn}\end{equation}
With the determinant (\ref{@fudetg1r}), the probability
becomes a constant
\begin{eqnarray}
P_0(0\, t_b|0\, t_a)=\const,
  \label{18.411fr1xp}
\end{eqnarray}
This shows that the retarded nature of the friction force
has
{\em subtracted\/} an energy
$\gamma/2$
from the energy in (\ref{18.411fr1x}).
With the ordinary path integral corresponding to a
Hamilton operator
with a
symmetrized term
 $-i(\hat p_v \hat v+
 \hat v
\hat p_v )/2$,
the subtraction of $ \gamma /2$  has changed this to
$-i \gamma \hat p_v \hat v$.

Note that the opposite case of an advanced
velocity term $\dot x^A$ in
(\ref{18.411fr})
would be approximated by a \ind{Drude function} $ \gamma^A_D(t)$
which looks just like
$ \gamma^R_D(t)$ in (\ref{@Drom}), but
with {\em negative\/} $ \omega _D$.
The right-hand side
of
 (\ref{@tracelogD})
becomes now $2\gamma$ rather than zero,
The corresponding formula for the functional determinant is
\begin{equation}
\Det (- \partial _t^2 -\gamma\partial_t^A) =
\exp\left[ \Tr\log(- \partial _t^2 -\gamma\partial_t^A)\right]
=\exp\left[(t_b-t_a) {\gamma} \right]
,
\label{@fudetg1a}\end{equation}
where $ \gamma \partial _t^A$
stands for
 the advanced  version of the functional matrix
(\ref{@retardedfn})
in which
$ \omega _D$
is replaced by
$ -\omega _D$.
Thus we find
\begin{eqnarray}
P_0(0\, t_b|0\, t_a)
& \propto&  \exp\left[ -(t_b-t_a) {\gamma}\right]  ,
  \label{18.411fr1xa}
\end{eqnarray}
with an {\em additional\/} energy $ \gamma /2$
with respect to the
ordinary formula
(\ref{18.411fr1x}).
This corresponds to
 the
 opposite (unphysical) operator order
$-i \gamma v \hat p_v $  in $\hat H_0$,
which would violate
the probability conservation of
time evolution twice as much as the symmetric order.

The above formulas for the
functional determinants can easily be extended to
the slightly more general case where $V(x)$ is the potential of a harmonic oscillator
$V(x)=M \omega _0^2x^2/2$.
Then the
path integral (\ref{18.411}) for the probability
 becomes
\begin{eqnarray}
P_0(x_b t_b|x_a t_a)
 \propto  \int {\cal D}x(t)\, \exp\left\{ -\frac{1}{2  w }
  \int_{t_a}^{t_b} dt\, [M\ddot x+M\gamma\dot x^R+ \omega _0^2x]^2\right\}\!,
  \label{18.411frt}
\end{eqnarray}
which we evaluate at $x_b=x_a=0$, where it is given by
the properly retarded expression
\begin{eqnarray}
\!\!\!\!\!\!\!\!P_0(0\, t_b|0\, t_a)
& \propto& \Det ^{-1}(- \partial _t^2
 -\gamma\partial_t+ \omega _0^2)
\nonumber \\
&\propto& \exp\left[-(t_b-t_a) \int _{-\infty}^\infty\frac{d \omega }{2\pi}
\log ( \omega'{} ^2-i \gamma  \omega' - \omega _0^2)\right] .
  \label{18.411fr1pw}
\end{eqnarray}
The roots of the argument of the logarithm lie now at $-i \omega _{12}$ with
\begin{equation}
 \omega _{12}=-\,\frac{ \gamma }{2}
\left(1\pm \sqrt{1-\frac{4 \omega _0^2}{ \gamma ^2}}\right).
\label{@}\end{equation}
Using the analytically regularized formula (\ref{@intanal1}),
the integral in (\ref{18.411fr1pw}) yields simply $ \gamma $,
and we obtain
the generalization of formula
(\ref{@fudetg1})
\begin{equation}
\Det (- \partial _t^2 -\gamma\partial_t- \omega _0^2)=
\exp\left[ \Tr\log(- \partial _t^2 -\gamma\partial_t- \omega _0^2)\right]
=\exp\left[(t_b-t_a) \frac{\gamma}2 \right].
\label{@fudetg1g}\end{equation}
The frequency
$ \omega _0$ has no influence upon the result.
This formula can be generalized further
to time-dependent coefficients
\begin{equation}
\Det \left[ - \partial _t^2 -\gamma(t)\partial_t- \Omega ^2(t)\right] =
\exp\left\{  \Tr\log\left[- \partial _t^2 -\gamma\partial_t- \Omega^2(t)\right]\right\}
=\exp\left[\int_{t_a}^{t_b} dt \frac{\gamma(t)}2 \right].
\label{@fudetg1gtd}\end{equation}
This follows from the factorization
\begin{equation}
\Det \left[ - \partial _t^2 -\gamma(t)\partial_t- \Omega ^2(t)\right] =
\Det \left[ \partial _t+ \Omega _1(t)\right]
\Det \left[ \partial _t+ \Omega _2(t)\right]
\label{@tdeppfo}\end{equation}
with
\begin{equation}
  \Omega _1(t)+
  \Omega _2(t)= \gamma (t),~~~~~\partial _t \Omega _2(t)+
  \Omega _1(t)
  \Omega _2(t)= \Omega ^2(t),
\label{@}\end{equation}
using the formula
\begin{equation}
\Det \left[ \partial _t+ \gamma (t)\right]=\exp\left[\frac{1}{2}
\int _{t_a}^{t_b}dt\, \gamma (t)\right] .
\label{2@intanal1t}\end{equation}
The probability
of the general path integral (\ref{18.411})
without retardation of the velocity term is therefore
\begin{eqnarray}
P_0(0\, t_b|0\, t_a)
& \propto&
 \exp\left[ -\frac{1}{2}(t_b-t_a)\gamma\right]  ,
  \label{18.411fr1x3}
\end{eqnarray}
as in (\ref{18.411fr1x}).

Let us now introduce retardation of the velocity term
by using the  $ \omega' $-dependent Drude expression (\ref{@Drom})
for the friction coefficient. First we consider again the harmonic
path integral (\ref{18.411frt}),
for which
 (\ref{18.411fr1pw}) becomes
\begin{eqnarray}
\!\!\!\!\!\!\!\!P_0(0\, t_b|0\, t_a)
\propto \exp\left\{-(t_b-t_a) \int _{-\infty}^\infty\frac{d \omega }{2\pi}
\log\left[  \omega'{} ^2-i \gamma^R_D( \omega )  \omega' - \omega _0^2\right] \right\} .  \label{18.411fr1ps}
\end{eqnarray}
For a large Drude frequency $ \omega _D\gg \gamma $
the roots are now
\begin{eqnarray}
 \omega _{12}=\frac{ \gamma }{2}\left(1\pm  \sqrt{1-\frac{4m_0^2}{ \gamma }} \right),
~~~~~~~
 \omega _{3}= \omega _D- \gamma .
\label{@}\end{eqnarray}
Using once more formula (\ref{@intanal1}), we see that $ \gamma $
 and the $m_0$-terms
disappear, and
we
remain with
\begin{eqnarray}
P_0(0\, t_b|0\, t_a)
&=&  \const.
  \label{18.411fr1x3b}
\end{eqnarray}
This implies a unit functional determinant
\begin{equation}
\Det( \partial  _t^2+i \gamma \partial_t^R+\omega _0^2)=1,
\label{@an V}\end{equation}
in contrast to the unretarded determinant
(\ref{@fudetg1g}).
By analogy with (\ref{@tdeppfo}), the also the general retarded determinant
is independent of $ \gamma (t)$ and $ \Omega (t)$.
\begin{equation}
\Det \left[ - \partial _t^2 -\gamma(t)\partial_t^R- \Omega ^2(t)\right] =1.
\label{@tdeppfotr}\end{equation}
In the advanced case, we would find similarly
\begin{equation}
\Det \left[ - \partial _t^2 -\gamma(t)\partial_{t}^A- \Omega ^2(t)\right] =
 \exp\left[ \int dt\,\gamma(t)\right]
.
\label{@tdeppfota}\end{equation}
All three determinants
are correct also for finite time intervals, due to the
solvability of the first-order differential equation
by means of an integrating factor.

\comment{The time order of the
potential $V'(x(t))$
with respect to the time slicing of the time derivatives  $\dot x $ and $\ddot x$
in the  time-sliced  Lagrangian (\ref{18.412})
is such that
\comment{chosen to agree
with
Eq.~(\ref{18.jac}) [to arrive at a constant Jacobi
determinant (\ref{18.deter})
and a\index{Langevin equation}
simple Langevin equation],
the potential is located in each
time slice at the initial time and the
Hamiltonian in Eq.~(\ref{18.420}).
The Hamilton operator in the Fokker-Planck equation
(\ref{18.424}) must have the
operators $
\hat p,
\hat p_v$, $ x$, and $v$
ordered correspondingly. Since $p_v$
stands
{\em before \/} $v\equiv \dot x$.
}}

Note that the retardation prescription can be avoided by
a trivial additive change of the Lagrangian
(\ref{18.412}) to
\begin{equation}
L_{\rm e}(x,\dot x) = \frac{1}{2w} \left [\ddot x+M \gamma \dot x +V'(x)\right ]^2
-\frac{ \gamma }{2}.
  \label{18.412ex}
\end{equation}
From this the path integral can be calculated
with the usual time slicing.

\section{Strong Damping}

For $ \gamma \gg V''(x)/M$,  the dynamics is dominated by dissipation,
and the Lagrangian (\ref{18.412}) takes a
more conventional form in which only $x$ and $\dot x$ appear:
\begin{equation}
L_{\rm e}(x,\dot x) =
\frac{1}{2w} \left [M\gamma\dot x^R+V'(x)\right ]^2=
\frac{1}{4D} \left [\dot x^R+\frac{1}{M \gamma }V'(x)\right ]^2,
  \label{18.413}
\end{equation}
where $\dot x^R$ lies slightly {\em before\/}
$V'(x(t))$.\index{overdamping}
The probability
\begin{equation}
P(x_b t_b|x_a t_a) = \int{\cal  D}x\, \exp\left[ -\int_{t_a}^{t_b} dt\,
L_{\rm e}(x,\dot x^R)\right]
  \label{18.413b}
\end{equation}
looks like an ordinary euclidean path integral for the density matrix
of a particle
of mass $M=1/2D$. As such it
obeys a differential equation of the Schr\"odinger type.
Forgetting a moment about the subtleties of the retardation,
we
introduce an auxiliary momentum integration and go over to
the canonical
representation of (\ref{18.413b}):
\begin{equation}
P(x_b t_b|x_a t_a) = \int {\cal D}x\, \int \frac{{\cal D}p}{2\pi}\,
  \exp\left\{ \int_{t_a}^{t_b} dt\,
  \left[ip\dot x-2D\frac{p^2}{2}+ip\frac{1}{M\gamma}V'(x)\right]
  \right\}.
  \label{18.414}
\end{equation}
This probability distribution satisfies therefore  the Schr\"odinger type
of equation
\begin{equation}
H(\hat p_b,x_b)P(x_b t_b|x_a t_a) = -\partial_{t_b}P(x_b t_b|x_a t_a)
  \label{18.415}
\end{equation}
with the Hamiltonian operator
\begin{equation}
H(\hat p,x) \equiv 2D\frac{\hat p^2}{2}-i\hat p\frac{1}{M\gamma}V'(x).
  \label{18.416}
\end{equation}
In order to conserve probability,
the momentum operator has to stand
to the left of
the potential term.
Only then does the integral over $x_b$
of Eq.~(\ref{18.415}) vanish.
\comment{Since the momentum variable is conjugate to $\dot x$,
it must have the same time argument as $\dot x$,
and  thus lies {\em after\/} $V'(x(t))$.
where the operator $\hat p$ stands
{\em after\/}  the operator $V'(x)$.}
Equation (\ref{18.415}) is the
overdamped
or
 ordinary
\index{Fokker-Planck equation}.

Without the retardation on $\dot x$ in
(\ref{18.413b}), the path integral
would certainly give a symmetrized
operator $-i[\hat p V'(x)+V'(x)\hat p]/2$ in $\hat H$.
This follows from the fact
that the
coupling
${(1/2DM\gamma )}\dot xV'(x)$ looks precisely like
the coupling of a particle
to a
magnetic field with
a ``\ind{vector potential}"{}
$A(x)=
{(1/2DM\gamma )}V'(x)$.
In this case we can also perform immediately the path integral
(\ref{18.413b})
\begin{eqnarray}
P_0(x_b t_b|x_a t_a)
 \propto  \int {\cal D}x(t)\, \exp\left\{ -\frac{1}{2  w }
  \int_{t_a}^{t_b} dt\, [M\gamma\dot x+V'(x)]^2\right\}\!
  \label{18.411frt0}
\end{eqnarray}
at $x_b=x_a=0$, which is given by
\begin{eqnarray}
\!\!\!\!\!\!\!\!P_0(0\, t_b|0\, t_a)
& \propto& \Det ^{-1}\left[
 \partial_t+ V'(x)/M \gamma \right] ,
  \label{18.411fr1p0q}
\end{eqnarray}
where from formula (\ref{2@intanal1t})
\begin{equation}
 \Det\left[
 \partial_t+ V''(x)/M \gamma  \right]
=     \exp\left[ \int dt \,V''(x)/2M \gamma \right] .
\label{@orres}\end{equation}

The effect of retardation of the velocity in (\ref{18.413})
is obvious since the trivial retarded determinant
(\ref{@tdeppfotr}) is independent of the strength of the damping:
\begin{equation}
 \Det\left[
 \partial_t^R+ V''(x)/M \gamma  \right] =1.
\label{@retresu}\end{equation}
In the advanced case one would have
\begin{equation}
 \Det\left[
 \partial_t^A+ V''(x)/M \gamma  \right]
=     \exp\left[ \int dt \,V''(x)/M \gamma \right] .
\label{@advresu}\end{equation}

For the differential equation (\ref{18.415}),
the difference between the ordinary and the retarded results
(\ref{@orres}) and (\ref{@retresu}) implies that
the initially symmetric operator order
 $-i[\hat p V'(x)+V'(x)\hat p]/2$ in $\hat H$
is changes into
 $
-i[\hat p V'(x)+V'(x)\hat p]/2 -V''(x)/2
-i\hat p V'(x)$,
as necessary for conservation of probability.

As in Eq.~(\ref{18.412ex}) we can avoid the
retardation of the velocity by
adding to the  Lagrangian
(\ref{18.413}) a term containing the second derivative of the potential:
\begin{equation}
L_{\rm e}(x,\dot x) = \frac{1}{4D} \left [\dot x+\frac{1}{M\gamma}V'(x)\right ]^2
-\frac{1}{2M \gamma}V''(x).
  \label{18.413ex}
\end{equation}
From this the path integral can be calculated
with the same slicing as for the gauge-invariant
coupling  of a magnetic vector potential
(see Sections 10.6 and 11.3 in the textbook \cite{PI})
\begin{eqnarray}
\!\!\!\!\!\!\!\!\!\!\!P_0(x_b t_b|x_a t_a)
 \propto  \int {\cal D}x(t)\, \exp\left[ -\frac{1}{4D }
  \int_{t_a}^{t_b} dt\,\left\{  \frac{1}{4D }\left[ \dot x+\frac{V'(x)}{M \gamma}
 \right] ^2-\frac{V''(x)}{2M \gamma }\right\}\!\right].
  \label{18.411frt0e}
\end{eqnarray}
\section{Langevin Equations}

For
for high $ \gamma T$.
the forward--backward path integral
(\ref{18.394})
has only small
fluctuations of $y$,
and
 $K(t,t')$
 becomes a $ \delta $-function.
Then we can expand
\begin{equation}
V\left(x+\frac{y}{2}\right)-V\left(x-\frac{y}{2}\right) \sim
  yV'(x)+\frac{y^3}{24}V'''(x)+\dots~,
  \label{18.395}
\end{equation}
keeping only the first term.
We further
introduce an auxiliary quantity $\eta (t)$ by
\begin{equation}
\eta (t)\equiv M\ddot x^R(t)+M\gamma\dot x(t)+V'(x(t)).
  \label{18.396}
\end{equation}
With this, the
exponential function in (\ref{18.394}) becomes
after a partial integration of the first term using the
endpoint properties
$y(t_b)=y(t_a)=0$:
\begin{equation}
\exp\left\{- \frac{i}{\hbar}
  \int_{t_a}^{t_b}
   dt\, y \eta - \frac{w}{2\hbar^2}
\int_{t_a}^{t_b}  dt\,
    y^2(t) \right\},
  \label{18.397}
\end{equation}
where
  $ w $ is the constant (\ref{18.400}).
The variable $y$ can obviously be integrated out and we find
a probability distribution
\begin{equation}
P[ \eta ]\propto \exp\left\{ -\frac{1}{2w}
  \int_{t_a}^{t_b}  dt\,
    \eta^2(t)  \right\},
  \label{18.398}
\end{equation}
The expectation value of an arbitrary
functional of $F[x]$
can be calculated from the path integral
\begin{eqnarray}
\langle F[x]\rangle_ \eta  \equiv {\cal N}
\int {\cal D}x\,  P[ \eta ]F[x],
\label{@Expecx}\end{eqnarray}
where the normalization factor ${\cal N}$ is fixed by the condition
$\langle \,1\,\rangle=1$.
By a change of integration variables from $x(t)$ to $ \eta (t)$,
the expectation value (\ref{@Expecx})  can be rewritten as a
functional integral
\begin{eqnarray}
\langle F[x]\rangle _ \eta \equiv {\cal N}
\int {\cal D} \eta \,  P[ \eta ]\, \,F[x],
\label{@Expece}\end{eqnarray}
In principle, the integrand contains
a factor $J^{-1}[x]$, where $J[x]$ is the functional Jacobian
\begin{equation} \label{18.deter}
J[x]\equiv\Det[\lfrac{  \delta  \eta (t) }{ \delta x(t')}]=
 \det[M\partial _t^2+M\gamma \partial_t^R+V''(x(t))].
 \label{18.fjac}
\end{equation}
However,  in Eq.~(\ref{@tdeppfotr}) we have seen that
the retarded determinant is
unity, thus justifying its omission
in (\ref{@Expece}).

The path integral
(\ref{@Expece})
may be interpreted as
an expectation value with respect to the solutions
of a \iem{stochastic differential equation}
(\ref{18.396}) driven by a
 Gaussian
random \iem{noise} variable $\eta(t) $
with a
correlation function
\begin{equation}
\langle \eta(t) \eta(t') \rangle_T =
w \,K(t-t'),
  \label{18.399}
\end{equation}
\comment{Let us check that the Jacobian is indeed
unity.
For this we go through the derivation of the influence functional
once more for a time-sliced initial path integral
(\ref{18.394}), the
{\em causality\/} of the dissipation
forces (the slowdown of motion sets in {\em after\/}
the friction force begins acting)
gives the relation (\ref{18.396}) the following order
of the slicing labels
\begin{equation} \label{18.jac}
\eta _n= \frac{M}{\epsilon }(x_n-2x_{n-1}+x_{n-2})
+\frac{M\gamma }{2}(x_n-x_{n-2})+\epsilon V'(x_{n-2}).
\end{equation}
Due to the positioning of $V(x)$ at the initial point $x_{n-2}$
of each time interval,
the determinant can easily be calculated recursively
starting from one end of the time axis
and proceeding to the other end.
The result is a constant
\begin{equation} \label{18.jac1}
J={\rm const,}
\end{equation}
Thus  the path integral over $x(t)$ in (\ref{@Expecx}) goes
directly over into the integral  over $\eta (t)$ in (\ref{@Expece}).}
The stochastic differential equation   (\ref{18.396})
has thus been derived
from the forward--backward path integral at
high temperatures.
\comment{Note that
a positioning of
$V(x)$ at the final point $x_n$
of each time interval
would also yield a constant but would
violate
causality.
To record the causality of the time slicing
in the continuum notation,
we shall
write the term $y\dot x $ in the path integral
(\ref{18.394}) as
$y\dot x^R $.}

In the overdamped limit,
the classical Langevin equation with inertia (\ref{18.396})
reduced to the
\iem{overdamped Langevin equation}:%
\footnote{%
The stochastic movement is now a
so-called {\iem{Wiener process}}.}
\begin{equation}
\eta (t)\equiv M\gamma\dot x(t)+V'(x(t)).
  \label{18.396x}
\end{equation}
The noise average is formed as in
(\ref{@Expece}).

\section{Supersymmetry}

Recalling the origin  (\ref{@orres}) of the extra last term
in the exponent of the path integral (\ref{18.411frt0e}),
this can  be
 rewritten in a slightly more implicit but useful way as
\begin{eqnarray}
&&\!\!\!\!\!\!\!\!\!\!\!\!\!\!\!\!\!\!P_0(x_b t_b|x_a t_a)
 \propto  \int {\cal D}x(t)\, \Det\left[ \partial _t+\frac{V''(x)}{M \gamma}
\right] \exp\left\{ -\frac{1}{4D }
  \int_{t_a}^{t_b} dt\,  \frac{1}{4D }\left[ \dot x+\frac{V'(x)}{M \gamma}
 \right] ^2\right\}    \nonumber \\&&
  \label{18.411frt0ee}
\end{eqnarray}
In the form (\ref{18.411frt0ee}),
the time ordering of the velocity  term
is arbitrary. It may be
quantum mechanical,
but equally well
retarded or  advanced, as long as
it appears in the same way in  both the
Lagrangian
and determinant.
An interesting structural observation is
possible by generating
the determinant
with the help of an
auxiliary  fermion field $c(t)$
from  a path integral over $c(t)$:
\begin{equation}
\det[ \partial _t+V''(x(t))/M \gamma ]\propto
\int {\cal D}c{\cal D}\bar c\,
e^{-\int dt\bar c(t)\left[ M\gamma \partial _t+V''(x(t))\right]  c(t)}.
\label{@jacobin2}\end{equation}
In quantum field theory, such auxiliary fermionic
fields are referred to as \iem{ghost fields}.
 With these we can  rewrite
the path integral (\ref{18.413b}) for the probability
as an ordinary path integral
\begin{equation}
P(x_b t_b|x_a t_a) = \int {\cal D}x \int {\cal D}c{\cal D}\bar c\,
 \exp\left\{  -
{\cal A}_{\rm PS}[x,c,\bar c]\right\} .
  \label{18.414x}
\end{equation}
where
${\cal A}_{\rm PS}$
is the euclidean action
\begin{equation}
{\cal A}_{\rm PS} =\frac{1}{2DM^2 \gamma ^2} \int _{t_a}^{t_b}dt
\left\{\frac{1}{2} \left [M \gamma \dot x+V'(x)\right ]^2
+\bar c(t)\left[ M\gamma \partial _t+V''(x(t))\right]  c(t)  \right\} ,
  \label{18.414xA}
\end{equation}
first written down  by Parisi and Sourlas.\footnote{\aut{G. Parisi} and \aut{N. Sourlas}, Phys. Rev. Lett. {\em 43}, 744
(1979); Nucl. Phys. B {\em206}, 321 (1982).}
This action has a particular property:
If we denote the expression in the first brackets by
\begin{equation}
U_x\equiv M \gamma  \partial _tx+V'(x),
\label{@}\end{equation}
the
 operator between the
Grassmann variables in (\ref{18.414xA}) is simply the functional derivative
of $U_x$:
\begin{equation}
U_{xy}\equiv \frac{ \delta U_x}{ \delta y}=M \gamma  \partial _t+V''(x).
\label{@}\end{equation}
Thus we may write
\begin{equation}
{\cal A}_{\rm PS} =\frac{1}{2D} \int _{t_a}^{t_b}dt
\left[
\frac{1}{2} U_x^2
+\bar c(t)  \, U_{xy}  c(t)  \right] ,
  \label{18.414xAA}
\end{equation}
where $U_{xy}c(t) $ is the usual short notation for
the functional matrix multiplication $\int dt'U_{xy}(t,t')c(t')$.
The relation between the two terms
makes this action
\iem{supersymmetric}. It is invariant
under transformations which mix the Fermi and Bose
degrees of freedom.
Denoting by $ \varepsilon $ a small
 anticommuting Grassmann variable,
 the
action is invariant under the field transformations
\begin{eqnarray}
 \delta x(t)&=&\bar\varepsilon c(t)+\bar c(t)\varepsilon,\\
 \delta \bar c(t)&=&-\bar\varepsilon U_x,~~~~~\\
 \delta  c(t)&=& U_x \varepsilon  .
\label{@sstra}\end{eqnarray}
The invariance
follows immediately
after observing that
\begin{equation}
 \delta U_x= \bar\varepsilon U_{xy} c(t)+\bar c(t) U_{xy} \varepsilon.
\label{@}\end{equation}

Formally, a similar construction
is also possible for a particle with inertia
in the path integral (\ref{18.411}),
which is an ordinary path integral involving the Lagrangian
(\ref{18.412ex}).
Here we can write
\begin{eqnarray}
\!\!\!\!\!\!\!\!P(x_b t_b|x_a t_a)
={\cal N}  \int {\cal D}x(t)\,J[x] \exp\left\{ -\frac{1}{2  w }
  \int_{t_a}^{t_b} dt\, [M\ddot x+M\gamma\dot x+V'(x)]^2\right\}\!.
  \label{18.411new}
\end{eqnarray}
where $J[x]$ abbreviates the  determinant
\begin{equation}
J[x]= \det[M\partial _t^2+M\gamma \partial _t+V''(x(t))]
\label{@}\end{equation}
which is known from formula (\ref{@fudetg1gtd}).
The path integral (\ref{18.411new}) is valid for {\em any\/} ordering of the velocity
term, as long as it is the same in the exponent
and the functional determinant.

We may now  express the functional determinant  as a
path integral over fermionic ghost fields
\begin{equation}
\!\!J[x]= \det[M\partial _t^2+M\gamma \partial _t+V''(x(t))]\propto
\int {\cal D}c{\cal D}\bar c\, e^{-\int dt\,\bar c(t) \left[
M\partial _t^2+M\gamma \partial _t+V''(x(t))\right]  c(t)},
\label{@jacobin1}\end{equation}
and rewrite the probability $P(x_b t_b|x_a t_a)$
as
an ordinary path integral
\begin{eqnarray}
&&\!\!\!\!\!\!\!\!\!\!\!\!\!\!\!
\!\!\!\!\!\!\!\!\!\!\!\!\!\!\!\!\!\!\!\!\!\!\!P(x_b t_b|x_a t_a)
 \propto  \int {\cal D}x(t)\int {\cal D}c{\cal D}\bar c
\exp\{-{\cal A}^{\rm KS}[x,\c,\bar c]\},
  \label{18.411x0}
\end{eqnarray}
  where ${\cal A}[x,\c,\bar c]$ is the euclidean action
\begin{eqnarray}
&&\!\!\!\!\!\!\!\!\!\!\!{\cal A}^{\rm KS}[x,\c,\bar c]\equiv
 \!\int_{t_a}^{t_b}\! dt\!\left\{
\frac{1}{2  w }
 [M\ddot x\!+\!M\gamma\dot x\!+\!V'(x)]^2
+\bar c(t)\left[
M\partial _t^2\!+\!M\gamma \partial _t\!+\!V''(x(t))\right]  c(t)\right\}.\nonumber \\
&&
  \label{18.411x}
\end{eqnarray}
This formal expression contains
subtleties arising from the
boundary conditions when calculating
the Jacobian  (\ref{@jacobin1})
from the
functional integral
on the right-hand side.
It is necessary to factorize the
second-order operator in the functional determinant and express
each factor as a determinant as in
(\ref{@jacobin1}).
At the end, the action is again supersymmetric, but there are twice as many
auxiliary Fermi fields \cite{KS}.

\section{Stochastic Quantum Liouville Equation}

At lower temperatures, where
 quantum fluctuations become important,
the forward--backward path integral
(\ref{18.394})
 does not allow us to
derive a Schr\"odinger-like differential equation
for
the probability distribution
$P(x \,v \,t|x_av_at_t)$.
To see the obstacle, we
go over to the
canonical representation
of (\ref{18.394}):
\begin{eqnarray}
&&\!\!\!\!\!\!\!\!\!\!\!\!|({x}_b t_b|{x}_a t_a)|^2 =
  \int {\cal D}x(t)\, {\cal D}y(t)\,
  \!\!\int \frac{{\cal D}p(t)}{2\pi}\,
   \frac{{\cal D}p_y(t)}{2\pi}\,
  \exp\bigg\{
\frac{i}{\hbar}
    \int \tab dt\, \left[p\dot x+p_y\dot y-H_T\right]
     \!\!  \bigg\}, \nonumber \\
&&
  \label{18.394neu}
\end{eqnarray}
where
\begin{equation}
 H_T=\frac{1}{M}p_yp_x+\gamma p_yy +V(x+y/2)-V(x-y/2) -
i\frac{ w }{2\hbar}y\hat{K}y
\label{canh}\end{equation}
plays the role of a Hamiltonian.
Here $\hat{K}y(t) $ denotes the
product of the functional matrix
$\hat{K}(t,t') $
with the functional vector
$y(t') $ defined by
$\hat{K}y(t) \equiv \int dt'K(t,t')y(t')$.
After omitting the $y$-integrations at the endpoints,
we obtain a path integral representation
for the product of
amplitudes
\begin{equation}
\!\! \rho (x_by_bt_b|x_ay_at_a)\equiv (x_b+y_b/2 ~t_b|x_a+y_a/2~ t_a)
(x_b-y_b/2 ~t_b|{x_a-y_a/2} ~t_a)^*.
\label{18.amplpr}\end{equation}
The Fourier transform of the diagonal elements of
this with respect to $y$ is the \ind{Wigner function}
\begin{equation}
W(x,p,t)\equiv \frac{1}{2\pi\hbar }
\int _{-\infty}^\infty
e^{ipy_b/ \hbar}
\rho (x\,y\,t\,|x\,y\,0).
\label{@}\end{equation}

When considering the change of
$\rho (x\,y\,t|x_ay_at_a)$ over a small time interval
$ \epsilon $,
the momentum variables
$p$ and $p_y$
have the same effect as differential operators $-i\partial_{x_b} $ and
$-i \partial  _{y_b}$, respectively.
The last term in $H_T$, however, is nonlocal in
time, thus preventing a derivation of a
Schr\"odinger-like
differential equation.

The locality problem can be removed
by introducing
a noise variable $ \eta (t)$
with the
correlations function
\begin{equation}
\langle   \eta (t) \eta (t')\rangle  _ \eta =\frac{ w }{2}K^{-1}(t,t').
\label{@averaging}\end{equation}
Then we can define a
temporally local
$ \eta $-dependent
Hamilton operator
\begin{equation}
\hat H_ \eta
\equiv \frac{1}{M}\left(\hat p_x+\gamma y\right)\hat p_y +V(x+y/2)-V(x-y/2) -
 y\eta
\label{canhn}\end{equation}
which governs
the evolution
of $ \eta $-dependent
versions of the amplitude products (\ref{18.amplpr})
via the \iem{stochastic Schr\"odinger equation}
\begin{equation}
i\hbar\partial _t
 \rho_ \eta (x\,y\,t|x_ay_at_a)
=\hat{H}_\eta  \,\rho_ \eta (x\,y\,t|x_ay_at_a).
\label{18.sla2}\end{equation}
Averaging this equation over $ \eta $  using
(\ref{@averaging})
yields for
$y_a=y_b=0$ the same probability distribution
as the
\ind{forward--backward path integral}
(\ref{18.394}):
\begin{equation}
|(x_bt_b|x_at_a)|^2 =
\rho (x_b\,0\,t_b|x_a\,0\,t_a)\equiv
 \langle\rho (x_b\,0\,t_b|x_a\,y_a\,t_a)\rangle_ \eta
\label{18.3.3}\end{equation}
At high temperatures, the noise averaged
stochastic Schr\"odinger equation
(\ref{18.sla2}) takes the form
\begin{eqnarray}
i\hbar \partial _t  \rho (x\,y\,t|x_a\,y_a\,t_a)
=\hat{\bar H}  \rho (x\,y\,t|x_a\,y_a\,t_a)
,
\label{18.timeeveq}\end{eqnarray}
where $\hat{\bar H}$ is the Hamiltonian associated
with the Lagrangian in the forward--backward path integral (\ref{18.394}):
\begin{equation}
\hat{\bar H}
\equiv \frac{1}{M}\hat p_y\hat p_x+\gamma y \hat p_y
+V(x+y/2)-V(x-y/2)
 -i\frac{w}{2 \hbar}y^2.
\label{canhn2}\end{equation}
In terms of the separate path positions $x_\pm=x\pm y/2$
where
$p_x=\partial _++\partial _-$ and $p_y=
(\partial _+-\partial _-)/2$,
this takes the
more
 familiar form
\begin{equation}
\!\!\!\!\hat{\bar H}
\equiv \frac{1}{2M}\left(\hat p_+^2-\hat p_-^2\right)
+V(x_+)-V(x_-)
+\frac{ \gamma }2
(x_+-x_-)
(\hat p_+ -\hat p_-)
 -i\hbar  \Lambda (x_+-x_-)^2.
\label{canhn2p}\end{equation}
In the last term we have introduced a useful quantity,
the so-called \iem{decoherence rate per square distance}
\begin{equation}
 \Lambda \equiv
\frac{w}{2\hbar ^2}=
\frac{M \gamma k_BT}{\hbar ^2}.
\label{@}\end{equation}
It is composed of the damping rate $ \gamma $ and
the squared thermal length
\begin{equation} \label{2.length}
l_\e(\hbar \beta )\equiv \sqrt{2\pi \hbar^2 \beta /M}
\end{equation}
as
\begin{equation}
 \Lambda =  \frac{2\pi \gamma }{l_\e^2(\hbar  \beta )},
\label{@}\end{equation}
and controls the rate of decay of interference peaks.\footnote{See
the collection of articles
\aut{D. Giulini}, \aut{E. Joos},
\aut{C. Kiefer},
\aut{J. Kupsch},
\aut{I.O. Stamatescu},
\aut{H.D. Zeh},
{\em Decoherence and the Appearance of a Classical World in Quantum Theory\/},
Springer, Berlin, 1996.}

Note that the order of the operators
in the mixed term of the form
$y\hat p_y
$
in Eq.~(\ref{canhn2})
is opposite to the mixed term $-i\hat p_v v$ in
the differential  operator (\ref{18.420})
of the Fokker-Planck equation.
This order is necessary to guarantee
the conservation of probability.
Indeed, multiplying the time evolution equation
(\ref{18.timeeveq})
by
$ \delta (y)$,
 and integrating both sides
over $x$ and $y$,
the left-hand side vanishes.

The correctness of this order can be verified
by calculating the fluctuation determinant
of the path integral
for the product of amplitudes
(\ref{18.amplpr})
in the Lagrangian form, which
looks just like
(\ref{18.394}), except that the
difference between forward and backward trajectories
$y(t)=x_+(t)-x_-(t)$
is nonzero at the endpoints.
For the fluctuation which vanish at the end points,
this is irrelevant.
As explained before, the order is a short-time issue, and we
can take $t_b-t_a\rightarrow \infty$.
Moreover, since the order is independent of the potential,
we may consider only the free case
$V(x\pm y/2)\equiv 0$.
The relevant fluctuation determinant
was calculated in
\comment{is [recalling the retarded-advanced structure of the exponent
implied by the influence functional (\ref{18.143a}) and formula (\ref{@fudetg1r})]
\begin{equation}
\Det
\left(
\begin{array}{cc}
0&\partial _t ^2+ \gamma^R _D \partial _t \\
\partial _t ^2+ \gamma^A _D \partial _t&w
\end{array}
\right)
=
\Det
\left(
\partial _t ^2+ \gamma^R _D \partial _t
\right)^2=1.
\label{@}\end{equation}
The unretarded determinant would have
yielded
$\exp\left[-\int dt \, \gamma (t)\right]  $, by
to
}
formula (\ref{@fudetg1}).
 In the Hamiltonian operator
(\ref{canhn2}), this implies an additional energy  $-i \gamma/2 $
with respect to the
symmetrically ordered term $
\gamma
 \{y,
\hat p_y\}/2
$,
 which brings it to
$\gamma
 y
\hat p_y
$,
 and thus the order in (\ref{canhn2p}).

\section{Conclusion}

With the help of analytic regularization
we have shown that the forward--backward
path integral of a point particle in a thermal bath of
harmonic oscillators
yields, at large temperature, a probability distribution
obeying a Fokker-Planck equation with the correct operator ordering
which ensures probability conservation.
By the same token, they yield the correct Langevin equations
with and without inertia.

\end{document}